\documentclass[twocolumn]{aastex7}

\newcommand{\fermi}{{\it Fermi}}
\newcommand{\target}{{PSR~J2032+4127}}
\usepackage{}

\begin{document}

\title{The orbital parameters of gamma-ray binary \target{}}

\author[0009-0005-2884-2941]{Yu-Feng~Luo}
\affiliation{School of Physics and Technology, Nanjing Normal University, Nanjing, 210023, Jiangsu, China}
\affiliation{Purple Mountain Observatory, Chinese Academy of Sciences, Nanjing 210008,  China}
\email{241002010@njnu.edu.cn}  

\author[0000-0001-7595-1458]{Shan-Shan Weng}
\affiliation{School of Physics and Technology, Nanjing Normal University, Nanjing, 210023, Jiangsu, China}
\affiliation{Nanjing key laboratory of particle physics and astrophysics, Nanjing, 210023, Jiangsu, China}
\email[show]{wengss@njnu.edu.cn}  

\author{Qing-Zhong~Liu}
\affiliation{Purple Mountain Observatory, Chinese Academy of Sciences, Nanjing 210008,  China}
\email[show]{qzliu@pmo.ac.cn}  

\author[0000-0002-2749-6638]{Ming-Yu Ge}
\affiliation{Key Laboratory of Particle Astrophysics, Institute of High Energy Physics, Chinese Academy of Sciences, Beijing 100049, China}
\email[show]{gemy@ihep.ac.cn}

\author[0009-0009-8477-8744]{Han-Long Peng}
\affiliation{School of Physics and Technology, Nanjing Normal University, Nanjing, 210023, Jiangsu, China}
\email{penghl@ihep.ac.cn}

\author[0000-0003-1480-2349]{Shi-Qi~Zhou}
\affiliation{School of Physics and Astronomy, China West Normal University, Nanchong 637002, People’s Republic of China}
\email{Pulsar.SqZhou@gmail.com}  

\author[0000-0002-0822-0337]{Shi-Jie~Gao}
\affiliation{School of Astronomy and Space Science, Nanjing University, Nanjing, 210023, China}
\email{gaosj@nju.edu.cn}  

\author{Yu-Jia~Zheng}
\affiliation{School of Physics and Technology, Nanjing Normal University, Nanjing, 210023, Jiangsu, China}
\email{yjzheng@ihep.ac.cn}  

\author{Yan~Zhang}
\affiliation{School of Physics and Technology, Nanjing Normal University, Nanjing, 210023, Jiangsu, China}
\email{1098631132@qq.com}  

\begin{abstract}

\target{} is the only one of gamma-ray binary, that exhibits pulsations in gamma-ray. Previous research has indicated that the pulsar and the Be star MT91 213 orbit each other in a highly eccentric orbit with an extremely long period, with the pulsar reaching its periastron on November 13, 2017. Since its launch, the \fermi{} satellite has been monitoring this pulsar for 16 years, covering the 8 years before and the 8 years after the pulsar passed its periastron. Using these data, we present an analysis of pulse arrival times, and precisely determine the orbital parameters for the first time: the orbital period of $P_{\rm orb} \sim 52.3$ yr, the eccentricity of $e \sim 0.98$, the semimajor axis of $a$sin$i \sim  25.3$ AU, and the orbital inclination of $\sim$ 47.1$^\circ$ -- 55.1$^\circ$. We also reveal another small glitch occurred in 2021, MJD $\sim$ 59500.

\end{abstract}


\section{Introduction} \label{sec:intro}
Gamma-ray binaries are a rare and enigmatic class of high-energy astrophysical systems consisting a compact object orbiting a massive, luminous companion star (typically an O or Be star). They emit the bulk of radiation in the gamma-ray regime ($>$ 1 MeV), making them among the most extreme particle accelerators.  The mechanisms behind gamma-ray emission is still under debate. There are two competitive models, i.e. the non-thermal emission originates from relativistic particles in the jet of microquasar, or in the shock due to the collision between the wind of the massive star and the pulsar wind. To date, radio pulsations have been detected in three systems, PSR~B1259-63 \citep{Johnston1992}, LS~I~+61$^{\circ}$~303 \citep{Weng2022}, and \target{}  \citep{Camilo2009}, supporting the latter scenario. Of greater interest,  only \target{} displays the gamma-ray pulsation \citep{Abdo2009}.

Historically, \target{} was discovered in the blind frequency searches using the Large Area Telescope (LAT) on the \fermi{}  {\it Gamma-ray Space Telescope} \citep{Abdo2009}.  Soon after that, its radio pulsation with the same period, $P = 143$ ms, was detected by the GBT. The measured period derivative of $\dot{P} = 2.0\times10^{-14}$ s~s$^{-1}$ implies a characteristic age $\tau = 110$ kyr, a characteristic magnetic field $B \sim 1.7\times10^{12}$ G, and a spin-down luminosity of $\dot{E} \sim 2.7\times10^{35}$ erg~s$^{-1}$  \citep{Camilo2009}. It was first thought to be an isolated pulsar, which powers the extended TeV source, TeV~J2032+4130 via its enigmatic pulsar wind \citep{Aliu2014}. But the subsequent radio observations revealed an anomalous change of spin-down rate, which was interpreted as the binary motion of the pulsar in a highly-eccentric wide orbit \citep{Lyne2015}.  Meanwhile, its optical companion is identified as a Be star MT91~213 lie in the Cyg OB2 stellar association at a {\it Gaia} distance of $\sim 1.7$ kpc \citep{Bailer-Jones2021}. The mass of  MT91~213 measured by different groups falls in a relatively narrow range, from 13.1 M$_{\odot}$ to 17.5 M$_{\odot}$ \citep{Kiminki2007, Wright2015, Ghoreyshi2024}.

\target{} passed through the periastron on 13 November 2017 \citep[MJD~58070.73,][]{Coe2017, Ho2017}. Extensive multi-wavelength observations were conducted to monitor the system before and after this epoch. While its optical, X-ray, and TeV emissions exhibited significant variations, the GeV emission remained stable \citep[e.g., ][]{Ho2017, Abeysekara2018, Ng2019, Chernyakova2020, Chen2022}. Orbital parameters are crucial for studying the radiation mechanisms and particle acceleration processes in gamma-ray binaries. Following \cite{Lyne2015}, \cite{Ho2017} updated the orbital period to 45--50 years by incorporating extended orbital phase coverage from radio and gamma-ray observations. However, the \fermi{}/LAT data collected during and after periastron passage have not been analyzed in the literature, which has prevented a precise determination of the orbital period. To date, \fermi{}/LAT has accumulated an additional eight years of data since the periastron epoch. In this work, we perform a timing analysis using the \fermi{}/LAT monitoring data with the aim of precisely determining the orbital parameters of \target{}.


\section{Observation and timing analysis} \label{sec:data}

The \fermi{}-LAT is a pair-production telescope that detects gamma rays by measuring electron-positron pairs, covering an energy range from 100 MeV to over 300 GeV with peak sensitivity at approximately 1 GeV \citep{Atwood2009}.
It consists of a tracker and a calorimeter, each of them made of a $4 \times 4$ array of modules, an anticoincidence detector that covers the tracker array, and a data acquisition system with a programmable trigger. Orbiting Earth every 90 minutes, \fermi{}-LAT scans the full sky approximately every three hours.  Taking advantage of its all-sky monitoring capability and long exposure times, we analyze \fermi{}-LAT observations of \target{} to constrain its binary orbital parameters.

\subsection{\rm Data reduction}
The gamma-ray flux from \target{} is stable, predominantly emitted within the 1--100 GeV range, and its SED peaks at $\sim$ 2.5 GeV. \citep{Chernyakova2020, Smith2023}. The data reduction and analysis were performed with Fermitools \citep[version 2.2.0., ][]{Atwood2009}. We selected Pass 8 Front + Back events \texttt{(evclass = 128 and evtype = 3)}\footnote{\url{https://fermi.gsfc.nasa.gov/ssc/data/analysis/scitools/binned_likelihood_tutorial.html}} within a $0.8^\circ$ circular region of interest (ROI), covering the energy range of 100 MeV – 20 GeV, where the pulsed emission is primarily detected, and the time interval from MJD 54682 to MJD 60949 (from 2008 August 4 to 2025 October 1). Events were filtered by a zenith angle cut of $< 90^\circ$ and high-quality events from good time intervals using the criteria \texttt{(DATA$\_$QUAL$>$0)$\&\&$(LAT$\_$CONFIG==1)}. Photon arrival times were corrected to the Solar-system barycenter using \texttt{gtbary} with the source position of R.A. = $308.054^\circ$  and Decl. = $41.457^\circ$.

\begin{table}
\caption{Parameters of PSR J2032+4127}
\label{table_timing_para}
\centering
\footnotesize  %
\begin{tabular}{l p{4cm} l}  %
\hline\hline
    &  Parameters              & Value                                                        \\
\hline
    & R.A.                     & $20^{h}32^{m}13^{s}.1247$                                      \\
    & Decl.                    & $+41^{\textordmasculine}$ 27$^{\prime}$24.3467$^{\prime\prime}$  \\
\hline
    & Epoch (MJD)                         & 58070                                              \\
    & $\it{\nu}$ (Hz)                         & 6.98086577 (1)                                 \\
    & $\it{\dot{\nu} }$ (10$^{-13}$ Hz $\rm s^{-1}$)  & -5.6636 (7)                                        \\
    & $\it{\ddot{\nu}}$ $(10^{-25}$ Hz $\rm s^{-2}$) & -1 (1)                                          \\
    &$\nu_{\rm 3}$ $(10^{-32}$ Hz $\rm s^{-3}$) &-1.0 (3)\\
    &$\nu_{\rm 4}$ $(10^{-40}$ Hz $\rm s^{-4}$) &1.8 (2)\\
    &$\nu_{\rm 5}$ $(10^{-48}$ Hz $\rm s^{-5}$) &1.7 (6)\\
    & $P_{\rm orb}$ (day)                            & 19111 (2)                                  \\
    & $\it{a}$sin$\it{i}$ (light-second)                & 12613.3 (9)                                       \\
    & \it{e}                                  & 0.979889 (2)                                       \\
    & $\omega{\textordmasculine}$        & 29.9680 (2)                                          \\
    & $T_{\omega}$ (MJD)                  & 58070.7351 (1)                                      \\
    & ${t_{\rm g1}}$ (MJD)                    &55810.76      \\
    & $\Delta \nu_{\rm g1}$ (10$^{-9}$ Hz)  &1908.9 (9)\\
    &$\Delta \phi_{\rm g1}$             &0.04 (2)\\
    &$\Delta \dot{\nu}_{\rm g1}$ (10$^{-15}$ Hz $\rm s^{-1}$)     &-0.63 (5)\\
    & ${t_{\rm g2}}$ (MJD)                   &59500            \\
    & $\Delta \nu_{\rm g2}$ (10$^{-9}$ Hz)  &251.7 (9)   \\
    &$\Delta \phi_{\rm g2}$            &0.14 (2)\\
    &$\Delta \dot{\nu}_{\rm g2}$ (10$^{-15}$ Hz $\rm s^{-1}$)    &-0.18 (3)\\
    &rms residuals (ms)    &2.032\\
\hline
\end{tabular}

\vspace{2mm}
\footnotesize
\noindent
Note: The $\Delta\phi$ denotes the phase offset,  while the subscripts $\mathrm{g1}$ and $\mathrm{g2}$ correspond to the parameters of the first and second glitches, respectively. The epoch of the first glitch ($t_{\mathrm{g1}}$) is adopted from \cite{Lyne2015}.
\end{table}


\begin{figure*}
\centering
\includegraphics[width=1\linewidth]{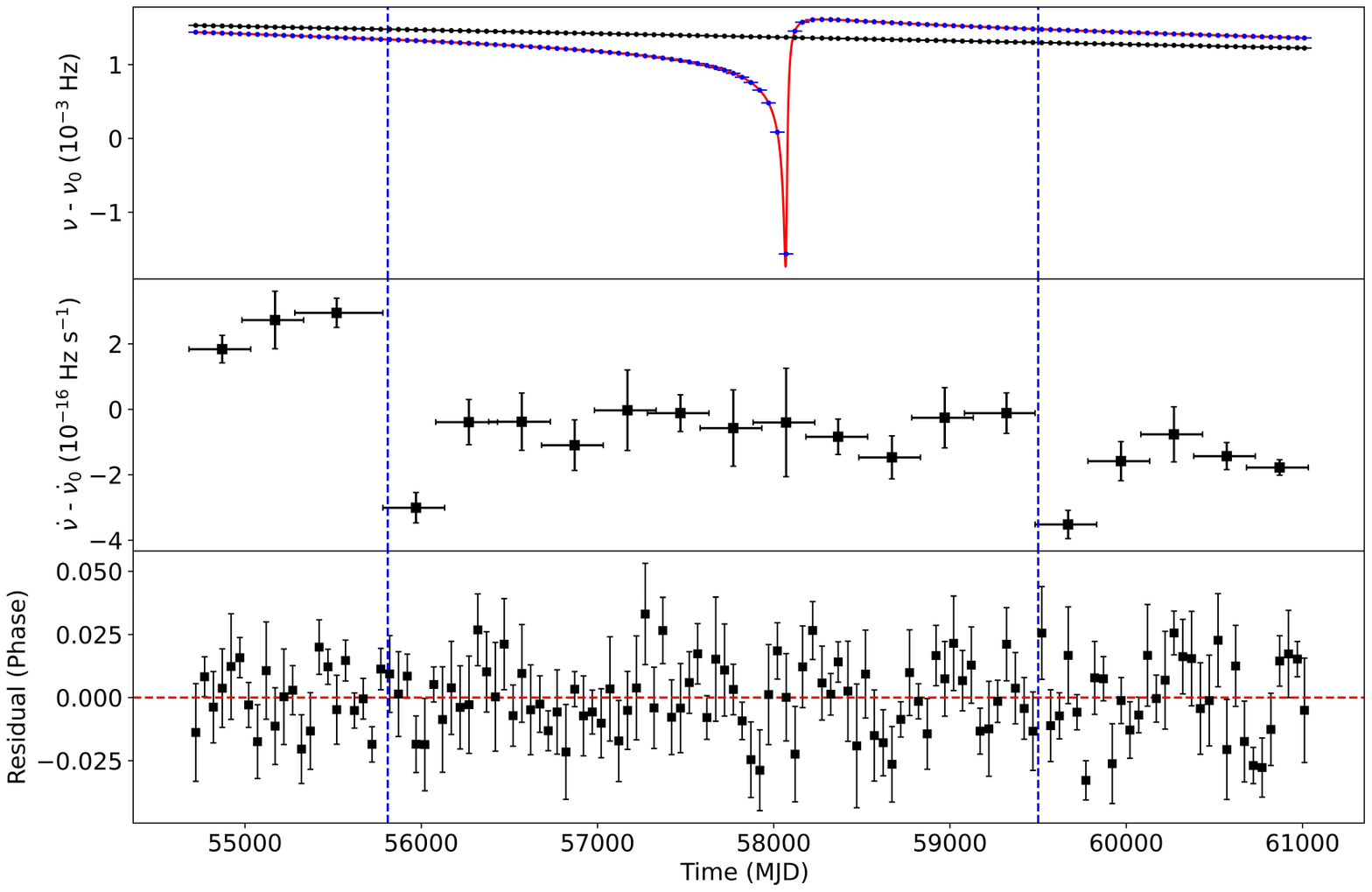}
\caption{The top panel illustrates the evolution of the neutron star's spin frequency $\nu$, with a reference value $\nu_0 = 6.9795$ Hz. Blue error bars denote the measured values, the red curve depicts the modeled spin frequency before orbital correction (derived from \texttt{TEMPO2}-fitted orbital and spin parameters), and black error bars show the intrinsic spin frequency after orbital demodulation. The middle panel shows the corresponding spin frequency derivative $\dot{\nu}$ ($\dot{\nu}_{0}$ = 5.6698 $\times$ 10$^{-13}$ Hz s$^{-1}$) as determined by \texttt{TEMPO2}. The bottom panel displays the fully coherent timing residuals (phase residuals) from the \texttt{TEMPO2} orbital fit.} \label{Spin_evolution}
\end{figure*}

\subsection{\rm Keplerian orbital parameter fitting}
The barycenter-corrected \fermi{} data (events file) were divided into consecutive 50-day segments. For each segment, we performed a pulsation search using photon arrival times, with the reference epoch set to the midpoint of the segment. This paper use the {\it Pearson} $\chi^{2}$ statistic \citep{Leahy1983,Leahy1987} for period search. A period search was conducted on the events data to derive $\nu$. To account for the substantial spin frequency derivative $\dot{\nu}$ near periastron, we extended the search to a two-dimensional $\chi^2$ grid over $\nu$ and $\dot{\nu}$. The resulting measurements are plotted as blue points in the top panel of Figure~\ref{Spin_evolution}. 
In contrast to the slow spin evolution typical of isolated neutron star (NS), the rapid evolution indicated by the blue data points must be attributed to modulation caused by binary orbital motion. To determine the intrinsic spin evolution of the NS, we first derived its five Keplerian parameters using Kepler Equation \citep{Damour1992}. The reference time for the 16-year dataset adopted in this work is MJD 58070. The resulting fit is plotted as the red curve in Figure \ref{Spin_evolution}. During the Kepler Equation fitting of Keplerian parameters and spin evolution by Python, we identified two glitches: one previously reported at MJD $\sim$ 55810.76 \citep{Lyne2015}, and another fixed at MJD $\sim$ 59500 in this work. The specific period jump will be discussed in Section \ref{Glitch}.

\subsection{\rm Barycentric correction for binary systems}
Using the five derived Keplerian orbital parameters, we applied a barycentric correction \citep{Riggio2008} to the photon arrival times of the NS, thereby referring them to the binary system barycenter. After all events data photons were corrected to the binary barycenter and search for the spin period was then performed on the corrected data in 50-day intervals, with the reference epoch for each interval taken as its midpoint. The results are shown as the black data points in the top panel of Figure \ref{Spin_evolution}.

\subsection{\rm TOAs of the binary system}
After the barycentric correction, we conducted a period search over the full dataset. The resulting spin evolution parameters were then used to phase-fold all photons and generate a standard pulse profile template. The data were subsequently divided into 50‑day segments, from which sub‑integrated pulse profiles were constructed. The TOAs were calculated using the relation:
${\rm TOA} = T_0 +P\times \Delta \Phi$
where $T_0$ is the start point of the data segment, ${P=1/{\nu}}$ is the pulse period, and $\Delta\Phi$ is the phase offsets. The phase offsets for the individual sub-integrated profiles were obtained by cross-correlating each profile with the standard template. The uncertainties in TOAs were estimated through 10,000 Monte Carlo simulations, where we generated multiple random profiles by introducing Poisson noise. At this point, we have obtained the barycentric-corrected TOAs. Subsequently, these TOAs were processed with the DD orbital model to reverse the binary barycentric correction, yielding the TOAs of the NS within the binary system.

\section{Results}
\subsection{\rm Orbital Period Determination with \texttt{TEMPO2}}
\begin{figure}[b]
\begin{center}
\includegraphics[scale=0.4]{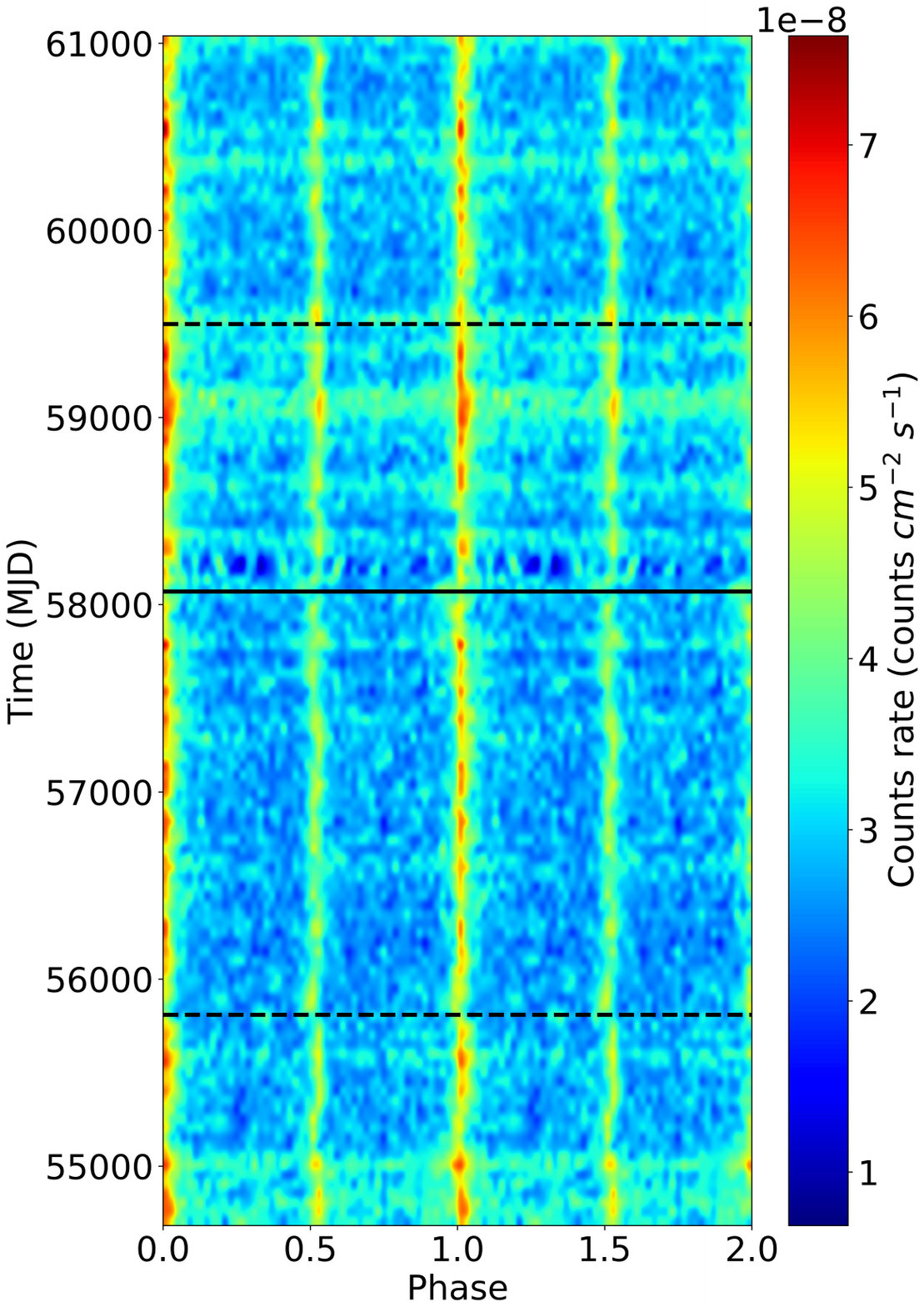}
\caption{The NS rotation phase-time diagram after the barycenter correction of the binary star, with the Epoch MJD is 58070. The black dashed lines mark two glitches at MJD ∼55810.76 and 59500, respectively. The black solid line indicates the time of periastron passage. All phases are aligned with the first segment's phase subtracted. \label{Time-PHI}}
\end{center}
\end{figure}

We therefore implemented a fully phase-coherent timing (FPCT) analysis using \texttt{TEMPO2}. The timing residuals are presented in the bottom panel of Figure \ref{Spin_evolution}, all measured TOAs were recorded in a timing file, while the five Keplerian parameters and spin evolution parameters were compiled into a parameter file for subsequent pulsar timing analysis, the results presented in this paper were obtained through iterative refinement and are summarized in Table \ref{table_timing_para}. Given to the 16-year span of the \fermi{} data, the timing solution involves a complex parameter set that makes it difficult to directly obtain accurate orbital parameters without proper initial values. Therefore, we initially employed partly phase-coherent timing (PPCT) to better constrain the initial orbital parameters and to study the local spin evolution while mitigating the effects of timing noise. In contrast, the PPCT first obtains one TOA per 50 days after barycentric correction of all photon arrival times. It then fits these TOAs in segments of seven (covering roughly 350 days) to derive a more accurate local ephemeris for each time span. The PPCT method is also designed to reduce timing noise and reveal the main characteristics of spin evolution, including identifying potential glitches, the time-phase diagram constructed from the phase-resolved 
PPCT results is shown in Figure \ref{Time-PHI}, with corresponding results also listed in Table \ref{Appendix tabel}. 

\begin{table*}
\caption{Partially phase-coherent timing parameter of PSR J2032+4127.}\label{Appendix tabel}
\begin{center} 
\rotatebox{90}{
\begin{tabular}{c c c c c c c c c}
\hline\hline
    &  MJD range     &Epoch         & $\nu$   &${\nu}_{\rm err}$       &$\dot{\nu}$   &${\dot{\nu}}_{\rm err}$      &$\ddot{\nu}$        &${\ddot{\nu}}_{\rm err}$                \\
\hline
& (MJD) & (Hz) & ($10^{-10}$~Hz) & ($10^{-13}$~Hz~s$^{-1}$) & ($10^{-17}$~Hz~s$^{-1}$) & ($10^{-23}$~Hz~s$^{-2}$) & ($10^{-23}$~Hz~s$^{-2}$) \\
\hline
&54682-55032 &54870.43 &6.9810223886 &4.27 &-5.6680 &4.21 &-1.15 &1.82\\
&54982-55332 &55170.35 &6.9810076990 &8.87 &-5.6671 &8.74 &1.91 &3.78\\
&55282-55782 &55520.33 &6.9809905606 &5.22 &-5.6669 &4.47 &0.51 &1.09\\
&55782-56132 &55970.31 &6.9809704321 &4.69 &-5.6728 &4.62 &-0.79 &2.00\\
&56082-56432 &56270.32 &6.9809557279 &7.49 &-5.6702 &6.90 &1.06 &3.12\\
&56382-56732 &56570.32 &6.9809410281 &8.93 &-5.6702 &8.75 &-1.43 &3.82\\
&56682-57032 &56870.31 &6.9809263291 &7.82 &-5.6709 &7.71 &2.49 &3.34\\
&56982-57332 &57170.30 &6.9809116320 &1.25 &-5.6698 &12.3 &3.13 &5.33\\
&57282-57632 &57470.31 &6.9808969359 &5.66 &-5.6699 &5.61 &-4.41 &2.41\\
&57582-57932 &57770.32 &6.9808822387 &11.8 &-5.6704 &11.7 &2.31 &5.05\\
&57882-58232 &58070.32 &6.9808675447 &16.6 &-5.6702 &16.6 &-14.0 &7.07\\
&58182-58532 &58365.26 &6.9808530953 &5.45 &-5.6706 &5.41 &-5.64 &2.32\\
&58482-58832 &58670.28 &6.9808381499 &6.65 &-5.6713 &6.56 &3.06 &2.84\\
&58782-59132 &58970.34 &6.9808234509 &9.34 &-5.6701 &9.20 &0.30 &3.99\\
&59082-59482 &59320.32 &6.9808063065 &5.25 &-5.6699 &6.17 &-3.01 &1.78\\
&59482-59832 &59670.32 &6.9807894097 &4.37 &-5.6733 &4.31 &-0.43 &1.87\\
&59782-60132 &59970.37 &6.9807747035 &6.06 &5.6714 &5.97 &0.20 &2.59\\
&60082-60432 &60270.30 &6.9807600078 &8.55 &-5.6706 &8.43 &1.61 &3.65\\
&60382-60732 &60570.30 &6.9807453113 &4.17 &-5.6712 &4.11 &-0.32 &1.78\\
&60682-60949 &60870.33 &6.9807306150 &2.29 &-5.6716 &2.36 &-2.10 &1.05\\
\hline
\end{tabular}}
\end{center}
\end{table*}

\subsection{\rm Glitch parameters}\label{Glitch}
While the spin evolution of a NS is typically stable, it can occasionally exhibit sudden jumps in spin rate, which are termed glitches. When fitting orbital parameters using TOAs in the \texttt{TEMPO2} software, apart from a previously detected glitch (at MJD 55810.76), the reference time for the first glitch in this article is fixed at this value, the phase calculation after the glitch is given by the following model \citep{Zhou2022}:
\begin{equation}\label{phase}
\Delta \phi(t) = \Delta \nu  \delta t+ \frac{1}{2}\Delta \dot{\nu}\delta t^2 + \Delta \nu_d \tau (1-e^{-\frac{\delta t}{\tau}})
\end{equation}
where $\delta t = t - t_g > 0$, $t_g$ is the time when the glitch occurs, the $\Delta \nu$ and $\Delta \dot{\nu}$ are the permanent changes in frequency and frequency derivative, the $\tau$ is the recovery timescale over which a transient frequency increment $\Delta \nu_d$ decays exponentially toward 0.37$\Delta \nu_d$  \citep[see e.g., ][]{Wong2001,Wang2012}. Since no clear exponential decay structure is seen in the residuals when the exponential term is included, we only fit for $\Delta \nu$ and $\Delta \dot{\nu}$, omitting the decay component. In addition, as noted earlier, our Python-based fitting indicates that besides the glitch near MJD$\sim$55810.76, a second, smaller-amplitude glitch occurs around MJD$\sim$59500. Both glitches are well described using the similar glitch parameters.

\subsection{\rm Orbital inclination}
\begin{figure}[t]
\begin{center}
\includegraphics[scale=0.3]{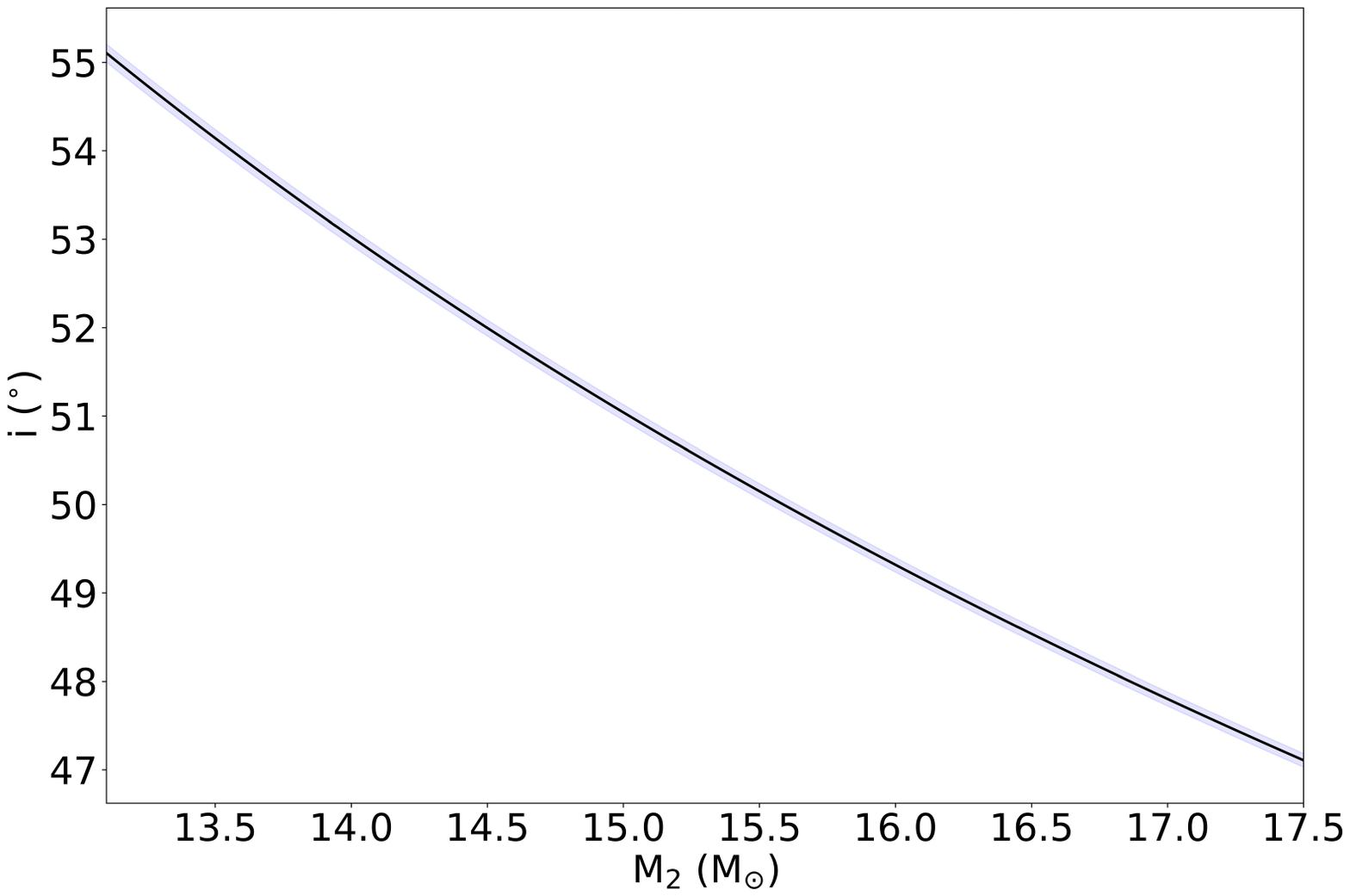}
\caption{The relationship between the orbital inclination and the Be star mass for a binary system with a NS primary of 1.4 M$_{\odot}$. The curve is derived from the mass function, showing the allowed combinations of inclination and Be star mass. The blue region indicates the uncertainty range.}\label{i_Mc}
\end{center}
\end{figure}

The mass of the Be star has been constrained by previous studies to the range of 13.1 M$_{\odot}$ -- 17.5 M$_{\odot}$ \citep{Kiminki2007, Wright2015, Ghoreyshi2024}. In this work, we calculated the mass function $f_m$ using the orbital parameters obtained from \texttt{TEMPO2}, and derived the relationship between the orbital inclination and the mass of the Be star based on Equation \ref{equ:7}:
\begin{equation}\label{equ:7}
f_{\rm m} \equiv \frac{(M_{\rm Be}\times \sin i)^3}{(M_{\rm NS} + M_{\rm Be})^2}=\frac{4{\pi}^2}{\rm G} \frac{(a{\rm sin}i)^3}{P_{\rm orb}^2}
\end{equation}
where, G is Newton’s gravitational constant, $M_{\rm NS}$ is the NS mass, fixed at the canonical value of 1.4 M$_{\odot}$, $M_{\rm Be}$ denotes the mass of the Be star, and $i$ is the orbital inclination of the binary system. The resulting dependence of the Be-star mass on the orbital inclination is presented in Figure \ref{i_Mc}.

\section{Summary}

The orbital parameters of \target{}—a long-period gamma-ray binary with stable pulsed gamma-ray emission—remain poorly constrained due to its long orbital period and high eccentricity. Based on the Doppler motion of the binary system, we fitted the orbital modulation of the observed pulse period and derived an orbital period of $P_{\rm orb}$ = 19110.5 $\pm$ 2.1 days, a projected semimajor axis $asin{\rm i}$ = 12613.2 $\pm$ 1.0 lt-s, and an eccentricity of $e$ = 0.979889 $\pm$ 0.000003. The wide orbital separation and the stable spin evolution suggest that the pulsar cannot enter either an accretion or a propeller state, but instead remains consistently in the rotation-powered regime, being similar to PSR~B1259-63 \citep{Shannon2014}. The precise orbital parameters derived in this work are of fundamental importance for calculating multi-wavelength radiation, including the spectral energy distribution and light curves.

In this work, the spin evolution from $Fermi$-LAT data was obtained through phase coherent timing analysis, yielding two small glitches. As listed in Table \ref{table_timing_para}, the parameters of the first glitch is consistent as reported by \cite{Lyne2015} and \cite{Ho2017} while the second one happened around MJD~59500 with $\Delta \nu_{g2} = 2.52(1) \times 10^{-7}$\,Hz, $\Delta \dot{\nu}_{g2} = -1.8(8) \times 10^{-16}\,\rm{Hz\,s^{-1}}$ as indicated by blue dashed lines in Figure \ref{Spin_evolution}.  For \target{}, the relative amplitudes ${\rm{\Delta \nu_{g}}/{\nu}}$ of these two glitches are $273.7(2) \times 10^{-9}$ and $36.1(2) \times 10^{-9}$, respectively. Both events are small glitches compared to the currently observed sample of over 216 pulsars that collectively exhibit hundreds ($>$ 790) of glitches \citep{Pagliaro2025,McKee2016,Serim2017,Zhou2022}. The glitch occurrence rates of \target{} is 0.12 yr$^{-1}$, which is significantly smaller than  Vela-like and Crab-like pulsars \citep{Espinoza2011}.  However, smaller glitches may remain undetected due to the lack of continuous radio monitoring data, such as Vela pulsar \citep{Espinoza2021}.

The mass function obtained from these results is $f_m=5.899 \pm 0.030$  M$_{\odot}$. Based on the Be star mass range of 13.1 M$_{\odot}$ $\sim$ 17.5 M$_{\odot}$ \citep{Kiminki2007, Wright2015, Ghoreyshi2024} established in previous work, the orbital inclination is constrained to $\sim$ 47.1$^\circ$ -- 55.1$^\circ$. This well-constrained inclination has important implications for understanding the system's geometry and potential observational characteristics.
\begin{acknowledgments}
We are very grateful to Tuan Yi for his useful comments. The authors thank supports from the National Natural Science Foundation of China under Grants 12473041, 12373051, 12473042, 12393812, 12233002 and 123B2045. This work is supported by the China Manned Space Program with grant no CMS-CSST-2025-A13. This work is supported in part by the National Key R\&D Program of China (Grant No. 2021YFA0718500 and 2023YFA1608100). This work is supported in part by China’s Space Origins Exploration Program.
\end{acknowledgments}

%
%

%

%
\facilities{\fermi{}-LAT}

\software{\texttt{Astropy} \citep{Astropy2013,Astropy2018,Astropy2022}, \texttt{TEMPO2} \citep{Hobbs2006}          }



\bibliography{psrj2032}{}
\bibliographystyle{aasjournalv7}



\end{document}